%% file: Yao_apj_Jan14.tex
\documentclass[onecolumn]{aastex631}
\usepackage{graphicx}% Include figure files
\usepackage{color}%for preprints
\usepackage[T1]{fontenc}
\usepackage{amssymb,amsmath}
\usepackage{ulem}
\usepackage{amsmath}
\usepackage{xcolor}
\usepackage{url}
\usepackage{textcomp}
\usepackage[titletoc]{appendix}
\usepackage{longtable}
\usepackage{subfiles}

\begin{document}

\title{Exploring the Neutrino Mass Hierarchy from Isotopic Ratios of Supernova Nucleosynthesis Products in Presolar Grains
}
\author{Xingqun Yao}
\email{sternyao@buaa.edu.cn}
\affiliation{School of Physics, 
Peng Huanwu Collaborative Center for Research and Education, and International Research Center for Big-Bang Cosmology and Element Genesis, Beihang University 37, Xueyuan Rd., Haidian-qu, Beijing 100191 China}
\affiliation{National Astronomical Observatory of Japan, 2-21-1 Osawa, Mitaka, Tokyo 181-8588, Japan}

\author{Toshitaka Kajino}
\email{kajino@buaa.edu.cn}
\affiliation{School of Physics, 
Peng Huanwu Collaborative Center for Research and Education, and International Research Center for Big-Bang Cosmology and Element Genesis, Beihang University 37, Xueyuan Rd., Haidian-qu, Beijing 100191 China}
\affiliation{National Astronomical Observatory of Japan, 2-21-1 Osawa, Mitaka, Tokyo 181-8588, Japan}
\affiliation{Center for Nuclear Study, The University of Tokyo, RIKEN campus, 2-1 Hirosawa, Wako, Saitama 351-0198, Japan}

\author{Yudong Luo}
\email{yudong.luo@pku.edu.cn}
\affiliation{School of Physics, Peking University, and Kavli Institute for Astronomy and Astrophysics, Peking University, Beijing 100871, P. R. China}

\author{Takehito Hayakawa}
\affiliation{Kansai Institute for Photon Science, National Institutes for Quantum Science and Technology, Kizugawa, Kyoto 619-0215, Japan}
\affiliation{Institute of Laser Engineering, Osaka University, Suita, Osaka 565-0871, Japan}

\author{Toshio Suzuki}
\affiliation{Department of Physics, College of Humanities and Sciences, Nihon University Sakurajosui 3-25-40, Setagaya-ku, Tokyo 156-8550, Japan}
\affiliation{NAT Research Center, NAT Corporation, 3129-45 Hibara Muramatsu, Tokai-mura, Naka-gun, Ibaraki 319-1112, Japan}

\author{Heamin Ko}
\affiliation{Institut für Kernphysik, Technische Universität Darmstadt, 64289 Darmstadt, Germany}
\author{Myung-Ki Cheoun}
\affiliation{Department of Physics and OMEG institute, Soongsil University, Seoul 07040, Korea}

\author{Seiya Hayakawa}
\affiliation{Center for Nuclear Study, The University of Tokyo, RIKEN campus, 2-1 Hirosawa, Wako, Saitama 351-0198, Japan}
\author{Hidetoshi Yamaguchi}
\affiliation{Center for Nuclear Study, The University of Tokyo, RIKEN campus, 2-1 Hirosawa, Wako, Saitama 351-0198, Japan}
\author{Silvio Cherubini}
\affiliation{Istituto Nazionale di Fisica Nucleare-Laboratori Nazionali del Sud, Via S. Sofia 62, I-95123 Catania, Italy}
\affiliation{Department of Physics and Astronomy, "Ettore Majorana" University of Catania, Via S. Sofia 64, I-95123 Catania, Italy}

\date{\today}
\begin{abstract}

We study the nucleosynthesis in a core-collapse supernova model including newly calculated neutrino-induced reaction rates with both collective and Mikheyev–Smirnov–Wolfenstein (MSW) neutrino-flavor oscillations considered. 
We show that the measurement of a pair of $^{11}$B/$^{10}$B and $^{138}$La/$^{139}$La or $^6$Li/$^7$Li
and $^{138}$La/$^{139}$La in 
presolar grains that are inferred to have originated from core-collapse supernovae 
could constrain the neutrino mass hierarchy. 
The new shell-model and the model of quasi-particle random phase approximation 
in the estimate of three important neutrino-induced reactions, $\nu+^{16}$O, $\nu+^{20}$Ne, and $\nu+^{138}$Ba are applied in our reaction network. 
The new rates decrease the calculated $^{7}$Li/$^{6}$Li ratio by a factor of five compared with the previous study. 
More interestingly, these new rates result in a clear separation of the isotopic ratio of $^{11}$B/$^{10}$B between normal and inverted mass hierarchies in the O/Ne, O/C, and C/He layers 
where $^{138}$La abundance depends strongly on the mass hierarchy.  
In these layers, the sensitivity of the calculated abundances of $^{10,11}$B and $^{6,7}$Li to the nuclear reaction uncertainties is also tiny. 
Therefore, we propose that the $^{11}$B/$^{10}$B vs. $^{138}$La/$^{139}$La and $^6$Li/$^7$Li vs. $^{138}$La/$^{139}$La 
in type X silicon carbide grains sampled material from C/He layer 
can be used as a new probe to constrain the neutrino mass hierarchy.

\end{abstract}
%\maketitle

\section{Introduction}

Core-collapse supernova (CCSN) ejects a huge number ($\sim 10^{58}$) \citep{Hirata:1987hu} of three-flavor neutrinos and their anti-neutrinos which provide observational signals 
for the collective neutrino flavor oscillation  \citep{Sigl:1992fn} and the Mikheyev–Smirnov–Wolfenstein (MSW) effect \citep{Wolfenstein:1977ue, Mikheev:1986gs}. 
Neutrinos launched from the surface of a proto-neutron star 
coherently scatter with one another and thereby change their flavors near the neutron-star atmosphere (collective oscillation). 
This effect induces an energy spectral split: the distribution at higher energies 
is swapped between $\nu_e$ and $\nu_{\mu,\tau}$ in the inverted hierarchy \citep{Pehlivan:2011hp}. 
In contrast, in the normal hierarchy, the energy spectra of the three-flavor neutrinos do not change remarkably. 
After the collective oscillation occurs, 
neutrinos propagate through the O/Ne layer and reach the O/C or C/He layer in which the electron number-density satisfies the MSW high-density resonance condition for the flavor change again \citep{Yoshida:2006sk, Yoshida:2008zb}. 
A substantial flavor conversion occurs between $\nu_e$ and $\nu_{\mu,\tau}$ for the normal hierarchy (${m_1<m_2<m_3}$)
and $\bar{\nu}_e$ and $\bar{\nu}_{\mu,\tau}$ for the inverted hierarchy (${m_3<m_1<m_2}$), respectively, 
where $m_i$ is the neutrino mass of each mass eigenstate \citep{Wolfenstein:1977ue}. 
In addition to these two effects of flavor conversions at high density, 
fast neutrino flavor conversion is also discussed in literature \citep{Xiong2020ApJ, Wu:2021uvt, Nagakura2022PhRvL}. 
Through these effects, 
the energy spectra of the three-flavor neutrinos exhibit specific variation at each position in the outer layers, 
contributing to the nucleosynthesis
which depends on the neutrino mass hierarchy. 

Experimentally, vacuum neutrino-oscillation project NO$\nu$A reported a result in favor of normal hierarchy at 1.96$\sigma$ C.L. \citep{NOvA:2019cyt, acero2022prd}, 
and combined Bayesian analysis of cosmological neutrino mass constraint prefers normal hierarchy \citep{Jimenez2022}.
Further studies of neutrino flavor conversion in vacuum and high densities are necessary both experimentally and theoretically to reach the final result of still unknown neutrino mass hierarchy. 

\cite{Woosley:1989bd} proposed that during the SN $\nu$-process 
neutrinos interact with abundant nuclei during propagation 
and lead to the synthesis of so-called $\nu$-isotopes such as $^{7}$Li, $^{11}$B, $^{19}$F, $^{92}$Nb, $^{98}$Tc, $^{138}$La, and $^{180}$Ta. 
This process has been extensively studied by many authors \citep{Heger2005PLB, Hayakawa:2010zzc,kobayashi2011ApJ, Hayakawa:2013ApJ, Hayakawa:2018ekx, Lahkar2017JApA, Sieverding2018rdt}. 
These works were improved by taking account of the flavor conversions at high density. 
Although the flavor change does not affect the neutral-current (NC) reactions 
like $^{16}$O$(\nu,\nu'n)^{15}$O, 
it impacts the final yields of 
$\nu$-isotopes through the charged-current (CC) reactions like $^{16}$O$(\nu_e,e^-p)^{15}$O, 
which depends strongly on the mass hierarchy \citep{Yoshida:2006qz, Yoshida:2008zb, Mathews:2011jq, Kusakabe:2019znq, Ko:2020rjq, Ko:2022uqv}, where charged and neutral current reactions are induced by the exchange of charged $W^{\pm}$ and neutral $Z^0$ bosons, respectively \citep{Weinberg:1995mt}. 
Therefore, the abundances of the synthesized isotopes at each layer reflect the neutrino mass hierarchy
through both the effects of collective and MSW flavor conversion. 
The predicted isotopic abundances exhibit significant differences between normal and inverted hierarchies 
when the collective oscillation effect maximally operates \citep{Ko:2020rjq, Ko:2022uqv}. 
It is also pointed out that other nucleosynthesis processes such as $\nu p$ and $\nu i$-process are affected by the collective neutrino oscillation \citep{sasaki_possible_2017, Balantekin2024ApJ}.

Presolar type X silicon carbide (SiC) grains, 
X grains, were identified in primitive meteorites based on their exotic isotopic compositions \citep{Amari1992ApJ, Nittler1996ApJ, Hoppe1996}. It is 
a group of presolar SiC grains with robustly inferred CCSN origins \citep{Liu2024SSRv}. 
The nucleosynthesis of Li and B isotopes in CCSNe could be probed by 
examining their isotopic compositions of X grains in primitive meteorites \citep{Hoppe2001}. 
Possible main production sources of the solar-system B isotope are the galactic cosmic ray (GCR) spallation reactions in space, 
CCSN neutrino process, and AGB stars; most of $^{10}$B originates from GCR, 
but $^{11}$B originates mainly from both GCR and CCSN \citep{Reeves1994}. 
X grains spent some time in the ISM before they were incorporated into the solar system so that their Li and B isotopic compositions could also have been additionally affected by GCR when they resided in the ISM. 
The previous theoretical studies \citep{Woosley95, Yoshida:2006qz, Austin2011, Prantzos2012AA} show that part of $^{11}$B in the solar system could arise from CCSNe nucleosynthesis. 
The boron and lithium isotopic ratios in the X grains have been measured,  $^{11}$B/$^{10}$B=4.68$\pm 0.31$ and $^7$Li/$^6$Li=11.83$\pm 0.29$ 
with 1$\sigma$ errors 
\citep{Hoppe2001, Fujiya_2011}, but these values agree with the solar values within $1.5\sigma$, 
$\rm (^{11}B/^{10}B)_{sol}$=4.04 and $\rm (^7Li/^6Li)_{sol}=$12.06 \citep{Lodders:2009}.
%The $^{11}$B/$^{10}$B and $^7$Li/$^6$Li 
%predicted in the standard GCR model are $\rm (^{11}B/^{10}B)_{GCR}$=2.5 and $\rm (^7Li/^6Li)_{GCR}=$1.4, respectively \citep{Reeves1994}. 
%They are lower than the values in the solar system, 
%small excess in boron isotopic ratio 
%suggests that part of $^{11}$B 
%in the presolar grain 
%The difference suggests that part of B and Li isotopic ratios \textbf{in the solar system 
%could arise from} CCSNe nucleosynthesis \citep{Woosley95, Yoshida:2006qz, Austin2011, Prantzos2012AA}.
On the other hand, $^{139}$La is mainly produced by s-process \citep{Bisterzo2014ApJ}, while $^{138}$La is regarded to be mainly produced via $\nu-$process \citep{Woosley:1989bd, Heger2005PLB, Hayakawa2008PhRvC}.
Therefore, 
if intrinsic $^{11}$B/$^{10}$B, $^{7}$Li/$^{6}$Li and $^{138}$La/$^{139}$La ratio can be obtained from X grains by suppressing asteroidal or terrestrial contamination, 
the data will provide valuable constraints on the astrophysical and physical parameters for the neutrino nucleosynthesis process. 

In the present work, we propose that 
correlated, intrinsic isotope ratios of $^{11}$B/$^{10}$B and $^7$Li/$^6$Li vs. $^{138}$La/$^{139}$La of presolar grains can 
directly constrain the neutrino mass hierarchy. 
We report here our theoretical study of the 
nucleosynthesis yields of $^6$Li, $^7$Li, $^{10}$B, $^{11}$B and $^{138}$La in a SN model 
by taking account of flavor conversions due to both collective and MSW effects. 
In addition to the $\nu+{^4}$He, $\nu+{^{12}}$C and $\nu+{^{138}}$Ba reaction rates, 
we use a series of newly calculated neutrino-induced reaction rates for $\nu+{^{16}}$O and $\nu+{^{20}}$Ne, 
which change the $^{6,7}$Li and $^{10,11}$B abundances drastically in the present calculations. 
We show that the two-dimensional plot of these isotopic ratios are sensitive probes of the mass hierarchy.

\section{Method}

\subsection{Supernova model and neutrinos}
The pre-SN model calculation has been performed based on the model for SN 1987A with metallicity $Z = Z_{\odot}/4$, 
the initial mass of the progenitor star is 20 $M_{\odot}$ \citep{Kikuchi:2015ena}. 
The star evolves to a helium core of 6 $M_{\odot}$ with the mass cut at $M_r=1.6\ M_{\odot}$, 
where $M_r$ is the Lagrangian mass coordinate. 
We adopted the hydrodynamic calculation for the explosion from \cite{Kusakabe:2019znq} and applied the resultant density and temperature profiles to the nucleosynthesis calculation. 
The hydrodynamic profile includes the region from $M_r=1.6M_{\odot}$ to $M_r=6.0M_{\odot}$ (helium core boundary), 
and is divided into 360 equal mass shells with 0.01224$M_{\odot}$. 

The total neutrino energy is $3 \times 10^{53}$ erg, 
and the exponential decay timescale of neutrino luminosity is set to be 3 s. 
We assumed that each neutrino flavor has the same luminosity (equal partition) and the energy spectra obey the Fermi-Dirac distributions with zero-chemical potentials. 
The initial temperatures are set to be 3.2, 5.0, and 6.0 MeV for $\nu_e$, 
$\bar{\nu}_e$, and $\nu_x$, respectively \citep{Yoshida:2008zb}. 
We adopted the same method as \cite{Ko:2022uqv} to calculate the flavor change due to the collective and MSW effects. 

\subsection{Neutrino induced reactions}\label{nu_induce}

In addition to the theoretical $\nu+^{12}$C and $\nu+^4$He reaction rates in \cite{Yoshida:2008zb}, 
we also calculated $\nu+^{16}$O and $\nu+^{20}$Ne reaction rates for various modes of final states using the shell model \citep{Suzuki:2018aey}. 
$^{138}$La is almost purely produced by the neutrino process via the reaction $^{138}$Ba$(\nu_e,e^-)^{138}$La, although a small amount of $^{138}$La comes from the $\gamma-$process $^{139}$La$(\gamma,n)^{138}$La as well \citep{Heger2005PLB}. 
We took the cross sections of this main reaction channel $^{138}$Ba$(\nu_e,e^-)^{138}$La 
from the quasi-particle random phase approximation (QRPA) calculation \citep{Cheoun:2010pm}. 
After a $\nu$-interaction populates excited states in a compound nucleus, 
the cross section leading to each final particle-emission channel is calculated using the branching ratios 
estimated in the Hauser-Feshbach statistical model \citep{Hauser1952} in the similar treatment as that in \cite{Suzuki:2018aey} and \cite{Cheoun:2010pm}. 
The other neutrino spallation reactions on A$\leq 70$ nuclei 
except for $^4$He, $^{12}$C, $^{16}$O and $^{20}$Ne 
are taken from \cite{Hoffmann:1992}.
The cross sections used in our nucleosynthesis calculations \citep{Cheoun:2010pm, Suzuki:2018aey} depend on the incident neutrino energy, 
which arises from the nuclear structure and available phase space. 
This is critically important because the energy-spectral change from the initial Fermi-Dirac distribution 
due to the flavor conversions at high density affects the production yields of $\nu$-nuclei via CC interactions, 
which depend on neutrino mass hierarchy. 
\cite{Hoffmann:1992} provides only the cross sections averaged over a normalized Fermi-Dirac distribution for a given initial temperature. 
We calculated the reaction rate by folding their averaged cross section multiplied by our modified neutrino-energy spectra after flavor conversion instead of Fermi-Dirac distribution. 
To clarify the effect of the updated $\nu+ ^{16}$O and $\nu+ ^{20}$Ne reaction rates, 
we compared the abundances calculated using 
both the reaction rates derived based on \cite{Suzuki:2018aey} (hereafter ``Su18'') and \cite{Hoffmann:1992} (hereafter ``HW92'') in the present study. 

\subsection{Uncertainties of Reaction Rates}
The flux-averaged measured $^{12}$C$(\nu_e,e^-)^{12}$N$_{g.s.}$ 
and $^{12}$C$(\nu,\nu')^{12}$C* reaction cross sections have $\pm10\%$ and $\pm20\%$ uncertainties 
for the CC and NC reaction channels, respectively \citep{KARMEN:1994xse}. 
The difference between the theoretical calculations of nuclear shell model \citep{Suzuki2006PRC, Suzuki:2018aey} 
and QRPA \citep{Cheoun:2010pm,cheoun_neutrinonucleus_2010} 
is less than the experimental uncertainties.
Thus, the uncertainties of the reaction rates for $^{12}$C and other target nuclei such as $^4$He, 
$^{16}$O, $^{20}$Ne and $^{138}$Ba 
are assumed to be $\pm10\%$ and $\pm20\%$ for the CC and NC reactions, respectively. 
We mentioned that the $^{11}$C$(\alpha,p)^{14}$N reaction rate 
has the largest uncertainty among all 91 nuclear reactions 
related to $^7$Li and $^{11}$B 
in the present calculations 
because this rate exhibits roughly one order of magnitude variation depending on poorly known resonance parameters at Gamow-window energy of 
$E_G$ = 0.25--1.09 MeV for effective temperature T = 0.2--0.8 GK of the SN $\nu$-process. 
This leads to the biggest uncertainty in the final $^{11}$B abundance. 

\section{Result and Discussions}

\subsection{Nuclear abundances and neutrino flavor conversions}

\begin{figure}[h]
	\begin{center}
		\includegraphics[width=0.8\textwidth]{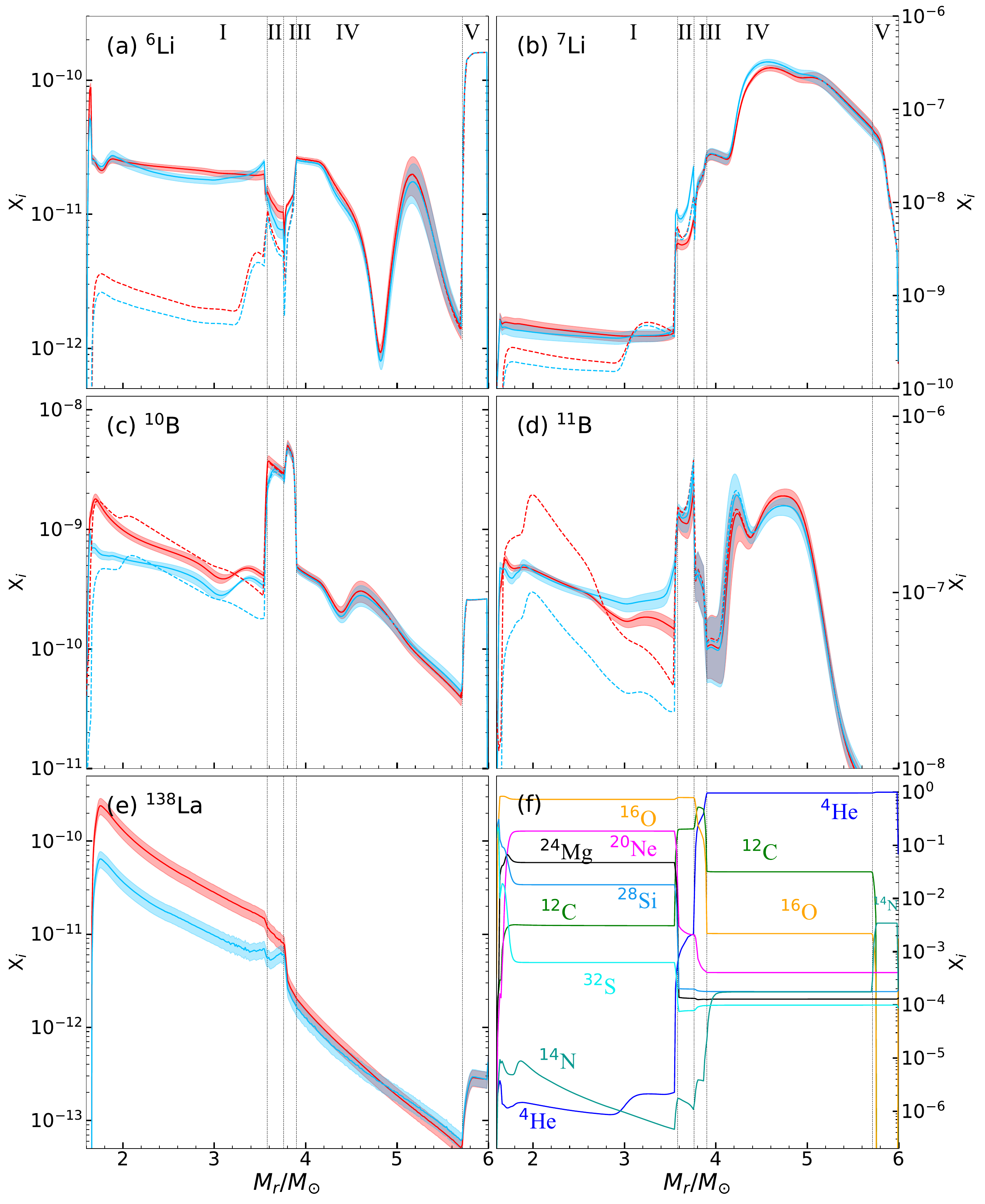}
		\caption{ Mass fractions of (a)$^6$Li, (b)$^7$Be+$^7$Li, (c)$^{10}$B+$^{10}$C, 
        (d)$^{11}$B+$^{11}$C, (e) $^{138}$La and (f) most abundant nuclei at 50 secs after the SN core-bounce as a function of the Lagrangian mass $M_r$ in the unit of $M_{\odot}$. 
		The red and blue colors in panels (a) to (e) represent the calculated results of the inverted and normal hierarchy cases, respectively. 
		Colored bands correspond to the uncertainties 
        arising from those of $\nu$-A reaction rates and $\rm ^{11}C(\alpha,p)^{14}$N reaction rate.  
		The dashed lines show the results by using the $^{16}$O+$\nu$ and $^{20}$Ne+$\nu$ rates from \cite{Hoffmann:1992} (HW92)
        instead of \cite{Suzuki:2018aey} (Su18). 
		$^{138}$La is mainly produced via $^{138}$Ba($\nu_e,e^-)^{138}$La. 
		We divide the star into five regions marked 
		\uppercase\expandafter{\romannumeral1} - \uppercase\expandafter{\romannumeral5} of O/Ne, O/C,
		C/He, He/C, and He/N layers, respectively. The C/He layer is the transitional zone between the O/C and He/C layers.
         %The five regions marked \uppercase\expandafter{\romannumeral1 -% \uppercase\expandafter{\romannumeral5} are O/Ne, O/C,
        %C/He, He/C, and He/N layers, respectively, 
        %following the scheme for 25 $M_{\odot}$ SN model \citep{meyer1995}.}
        Note that the scales of the vertical axis are different for each panel.}
		\label{c11_abun}
	\end{center}
\end{figure}

Figure \ref{c11_abun} shows the mass fractions of $\rm (a) ^6Li,
\ (b)^7Li+^7$Be (decay to $^7$Li with t$_{1/2}=$53 days), 
(c)$^{10}$B+$^{10}$C (decay to $^{10}$B with t$_{1/2}=$19.3 secs)\footnote{Notice that, in principle, we should consider $^{10}$Be contribution here since it has a lifetime t$_{1/2}=1.5\times10^{6}$ yrs. 
However, the yield of $^{10}$Be in our calculation is 15 orders of magnitude lower than $^{10}$B (see Table \ref{nuclei_table}), and we can ignore this yield contribution.}, 
(d)$\rm {^{11}B}+{^{11}C}$ (decay to $^{11}$B with t$_{1/2}=$20 mins), 
(e)$^{138}$La and (f) several abundant nuclei after the explosion as a function of $M_r$, 
where only the stable nuclei are marked in each panel, e.g. (b) $^7$Li 
represents $^7$Li+$^7$Be, etc. 
The red and blue solid lines with bands represent the calculated results in the inverted and normal hierarchies, respectively. 
The band width arises from the  $\nu$-A and $\rm ^{11}C(\alpha,p)^{14}$N reaction rate uncertainties discussed above. 
The calculated abundances with HW92 calculation are plotted as dashed lines. 
%The overall trend of the two results is similar with some exceptions. 

We divide the star into five regions marked 
\uppercase\expandafter{\romannumeral1} - \uppercase\expandafter{\romannumeral5} of O/Ne, O/C,
C/He, He/C, and He/N layers, respectively, following the scheme for 25 $M_{\odot}$ SN model \citep{meyer1995}, 
although region III is a transitional layer between the O/C and He/C layers. 
In Fig. \ref{c11_abun} (a), the Su18 calculated results are significantly higher than the HW92 calculations by one order of magnitude in the O/Ne layer (region I). 
This is due to the production of $^6$Li via the $\nu+ ^{16}$O reaction, which was not considered in the previous nucleosynthesis calculations (\cite{Ko:2022uqv} and references therein), although 
$^{16}$O is the most abundant nucleus in this region, as displayed in Fig. \ref{c11_abun} (f).  
The $^6$Li abundance in the inverted case (red) is slightly higher than in the normal case (blue). 
Similarly, the additional $^7$Li is produced via the $\nu+ ^{16}$O reaction in O/Ne layer of Fig. \ref{c11_abun} (b). 
However, in the C/He (region III) and He/C layers (region IV), 
the $^{6,7}$Li are mainly produced by other reactions such as $\rm ^3H(\alpha,\gamma)^7Li$, 
so that the solid lines and dashed lines are identical.

\begin{figure}
	\begin{center}
		\includegraphics[width=8.0cm]{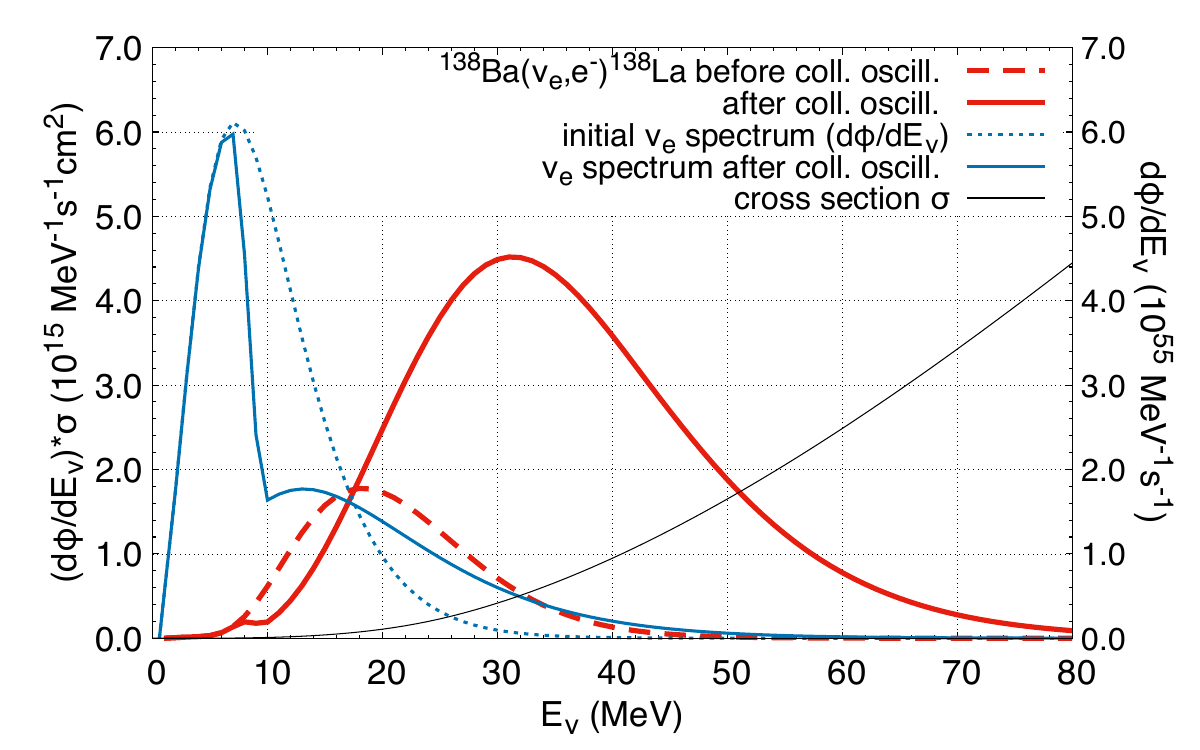}
		\caption{The dotted and solid blue lines are the electron-neutrino differential energy-spectra, $4\pi r^2 d\phi_{\nu}/dE_{\nu}$, 
        at 10 km and 1000 km which are respectively before and after the collective oscillation. 
        The black thin line shows the cross section of the $\rm ^{138}Ba(\nu_e,e^{-})^{138}La$ reaction from \cite{Cheoun:2010pm}. 
        Note that the cross-section is shown such that the end-point at 80 MeV is set to be 8.894$\rm \times 10^{-39}\ cm^2$. 
        The dashed and solid red lines present the electron neutrino-spectra multiplied by the $\rm ^{138}Ba(\nu_e,e^{-})^{138}La$ reaction cross section, $4\pi r^2 d\phi_{\nu}/dE_{\nu}\times \sigma$, 
        before and after the collective oscillation. 
		}
		\label{spectr}
	\end{center}
\end{figure}

In Fig. \ref{c11_abun} (c) and (d), most of $^{10,11}$B are produced in the O/C layer (region II) by the $\nu+^{12}$C reaction 
because there are plenty of $^{12}$C (see Fig.\ref{c11_abun} (f)). 
However, in C/He and He/C layers (region III and IV), $^{11}$B is not particularly enhanced because of the secondary destruction reaction of $^{11}$C$(\alpha,p)^{14}$N 
induced by abundant $^4$He. 
%via the $\rm ^7Li(\alpha,\gamma)^{11}B$ and $\rm ^7Be(\alpha,\gamma)^{11}C$ reactions. 
%Furthermore, the $^{16}$O$+\nu$ reaction produced additional $\alpha$ and proton particles 
In the O/Ne layer (region I) of Fig. \ref{c11_abun} (d), 
there is a difference in the $^{11}$B abundances between the solid and dashed lines 
using the two different sets of reaction cross sections for $\nu+^{16}$O and $\nu+^{20}$Ne, 
as described in subsection \ref{nu_induce}. 
$^{11}$B abundances by using the previous cross section data (red and blue dashed lines) 
decrease as $M_r$ increases because the neutrino-flux is a decreasing function of $M_r$ \citep{Kusakabe:2019znq}. 
Moreover, the abundance of $^{11}$B in the inverted hierarchy using the previous data (red dashed line) mostly comes from the $^{11}$C produced by the
$^{12}$C$(\nu_e,e^- p)^{11}$C reaction \citep{Ko:2022uqv} because the high-energy parts of the spectra of $\nu_e$ and $\nu_{\mu,\tau}$ are exchanged due to the collective oscillation \citep{Pehlivan:2011hp}, 
namely the $\nu_e$ spectrum is much enhanced than $\bar{\nu}_e$ (see blue line in Fig. \ref{spectr}). 
After adding the energy dependent $\nu+^{16}$O and $\nu+^{20}$Ne cross sections (solid lines), 
the final $^{11}$B abundance changes drastically such that the large difference of the dashed lines between the two hierarchies almost disappears. 
%These dashed lines are calculated using energy-independent cross sections from \cite{Hoffmann:1992}. 
%It is because $^{12}$C+$\nu$ cross-section is assumed to be energy independent, %which causes an unrealistic $^{11}$B yield when the cross section is folded by the modified neutrino differential energy-spectra due to the flavor conversions, 
%As the incident energy of neutrino increases, 
%the available number of final states for all possible transitions and phase space increases. 
%The $\nu$-induced cross sections are the increasing functions of energy as calculated by \cite{Suzuki:2018aey} and \cite{Cheoun:2010pm}. 
This is because the $\nu+^{16}$O reaction produces additional $\alpha$ particles that destroy $^{11}$C by the $^{11}$C$(\alpha,p)^{14}$N reaction, 
where $^{11}$C was produced from the $^{12}$C$(\nu_e,e^-p)^{11}$C reaction. 
%most strongly among all possible light-mass nuclear reactions. 
The resultant $M_r$-dependence of $^{11}$B is smoothed out as shown in Fig. \ref{c11_abun}(d).

The neutrino flavor-conversion due to the MSW resonance effect near 
C/He and He/C layers almost offsets the spectral change of the neutrino energy-distributions arising from the collective oscillation. 
In He-rich layers, spallation products of the $\nu+^4$He reaction, i.e. p, n, d, $^3$H and $^3$He, 
and the abundant $\alpha$ particles enhance both production and destruction of A=7 and 11 nuclei via $^3$H$(\alpha,\gamma)^7$Li$(\alpha,\gamma)^{11}$B$(\alpha,n)^{14}$N and $^3$He$(\alpha,\gamma)^{7}$Be$(\alpha,\gamma)^{11}$C$(\alpha,p)^{14}$N. 
Therefore, 
one cannot find any appreciable difference in any nuclei between the two hierarchies within the reaction uncertainties in He/C layer, 
especially in the outermost region $M_r\geq 5 M_{\odot}$. 
%The resultant red and blue for $^7$Li and $^{11}$B become close to each other. 
%Therefore, the red and blue bands are more close to each other than 
%the dashed lines in the region \uppercase\expandafter{\romannumeral1} in Fig. \ref{c11_abun} (d). 
%The most abundant nucleus distribution can be referred to in Fig. \ref{c11_abun} (f). 
%And $^{12}$C$(\nu_e,e^- p)^{11}$C rate is enhanced by the SI effect in the inverted hierarchy case, 
%In region \uppercase\expandafter{\romannumeral4}, 
%there is no difference between dashed and solid lines 
%due to the low $^{16}$O and $^{20}$Ne abundance in He-layer.

%\subsection{The role of neutrino flavor change effect}

As for $^{138}$La in Fig. \ref{c11_abun} (e), 
there is a noticeable disparity between the normal and inverted hierarchies in O/Ne layer (region I). 
The mechanism is similar to that for $^{11}$B in Fig.\ref{c11_abun} (d), 
except that $^{138}$La production is negligible from the $\nu+^{16}$O and $\nu+^{20}$Ne reactions 
because $^{138}$La is so heavy that the light-mass charged particles have tiny cross sections due to the Coulomb barrier in the secondary destruction process, 
except for neutron-induced reactions which have already been included in our network calculations. 
As shown in Fig. \ref{spectr}, 
the modified $\nu_e$ energy spectrum (blue solid line 4$\pi r^2 \frac{d\phi_{\nu_e}}{dE}$) is larger than the initial one (blue dotted line) above $~$18 MeV due to the collective oscillation. 
Therefore, the production rate of $^{138}\rm{La} \propto 4\pi r^2\frac{d\phi_{\nu_e}}{dE}\times \sigma$ (red solid line) is much larger than the initial value. 
In the outer regions of the C/He and He/C layers (region III and IV), 
the MSW effect almost offsets the collective effect, 
leaving only a tiny difference between the two hierarchies, similar to $^{6,7}$Li and $^{10,11}$B. 
The difference of $^{138}$La is sensitive to the neutrino mass hierarchy only in O/Ne layer and O/C layer (region I and II). 
These mechanisms are valid under the condition $T_{\nu_e}<T_{\nu_x}$ and $T_{\bar{\nu}_e}<T_{\nu_x}$, 
and the result of relative abundance ratios is less sensitive to the adopted SN model. 

To summarize, the most important reactions in O/Ne layer (region I) for B and Li production are 
neutrino-induced spallation reactions $\nu+^{12}$C and $^{16}$O. 
While in C/He and He/C layers (region III and IV), the most significant reactions are $(\alpha,\gamma)$ and $(\alpha,p/n)$ reactions, for the production and destruction, respectively. 
As for $^{138}$La, it is mainly produced from the $^{138}$Ba$(\nu_e,e^-)^{138}$La reaction in the O/Ne layer. 
Photo-induced reactions do not operate efficiently because photon number density above the threshold energy for the production of these isotopes is extremely small with the Planck distribution of photons at the effective temperature T$\simeq 0.1-1$ GK for explosive SN nucleosynthesis. 

Based on these nucleosynthesis calculations, the isotopic productions $\Delta M(^A$Z) and overproduction factor $\Theta (^A$Z) are listed in the appendix Table \ref{nuclei_table} 
in the same tabular format of \cite{meyer1995}. 
In this table, all values are taken from our CCSN model 50 secs after the explosion, 
therefore, some radioactive nuclei appear in the table. 
As mentioned above, we divided the model into five major layers: O/Ne, O/C, C/He, He/C and He/N layers. 
To show the very inner structure of our model, 
we also divided the inner part ($M_r<1.71M_{\odot}$) into three zones as Fe/Ni, Si/S, and O/Si zones, depending on the most abundant nuclei of each zone. 
The overproduction factor $\Theta (^A$Z) is the ratio of $^A$Z abundance to the corresponding solar abundance, 
which is the same as in \cite{meyer1995}. 
For the radioactive nuclei, $\Theta (^A$Z) was calculated with the corresponding final stable daughter nucleus in solar system.

%Therefore, the $^7$Li/$^6$Li, $^{11}$B/$^{10}$B and 
%$^{138}$La/$^{139}$La abundance ratios in presolar grains originating 
%from SNe, thus, could constrain the conditions of the neutrino mass 
%hierarchy and the degree of mixing between layers from which SN grains are formed in an outflow ejecta. 

\subsection{Isotopic ratios and X grains}

\begin{figure}
    \centering
    \includegraphics[width=0.4\linewidth]{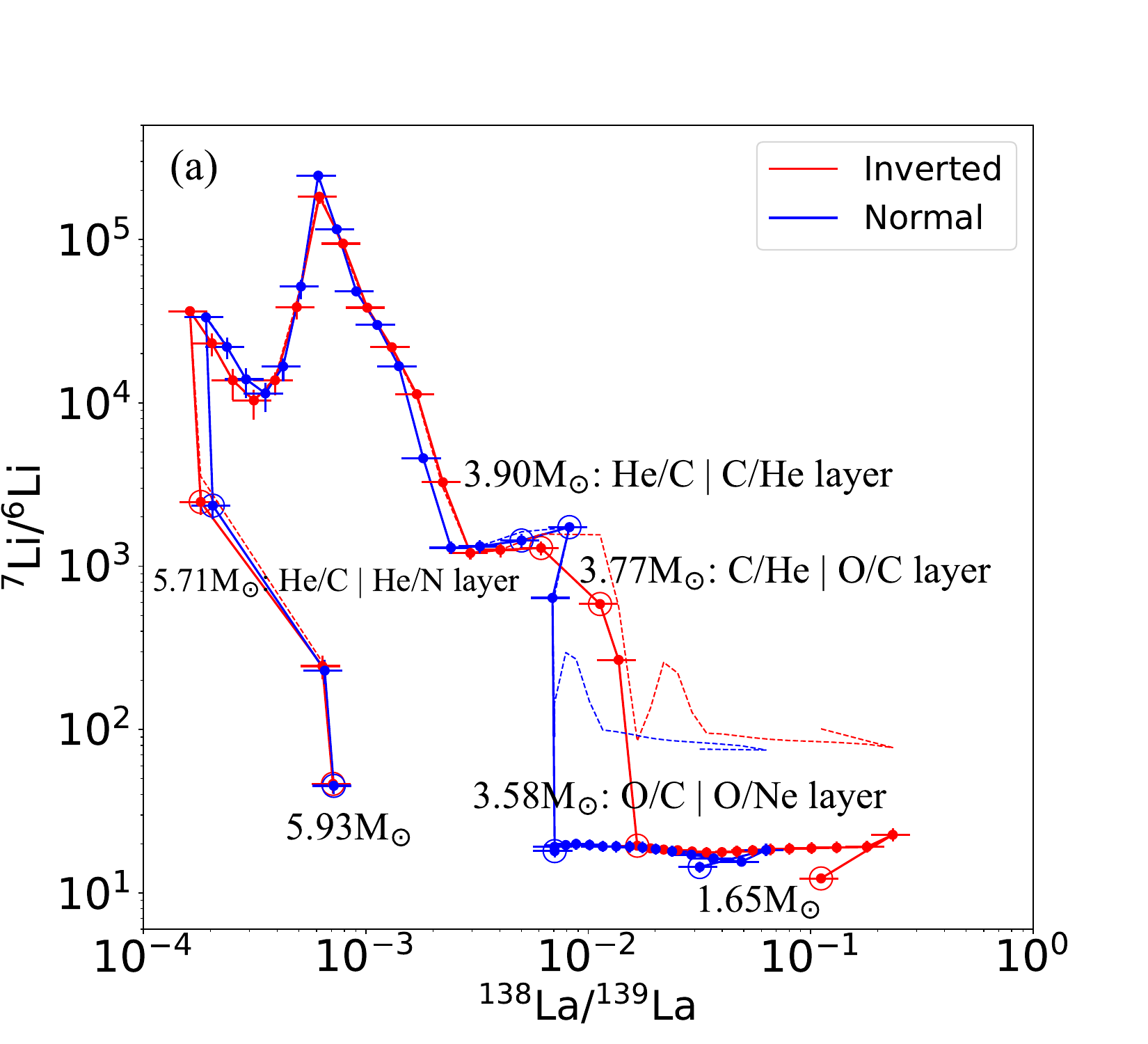}
    \includegraphics[width=0.4\linewidth]{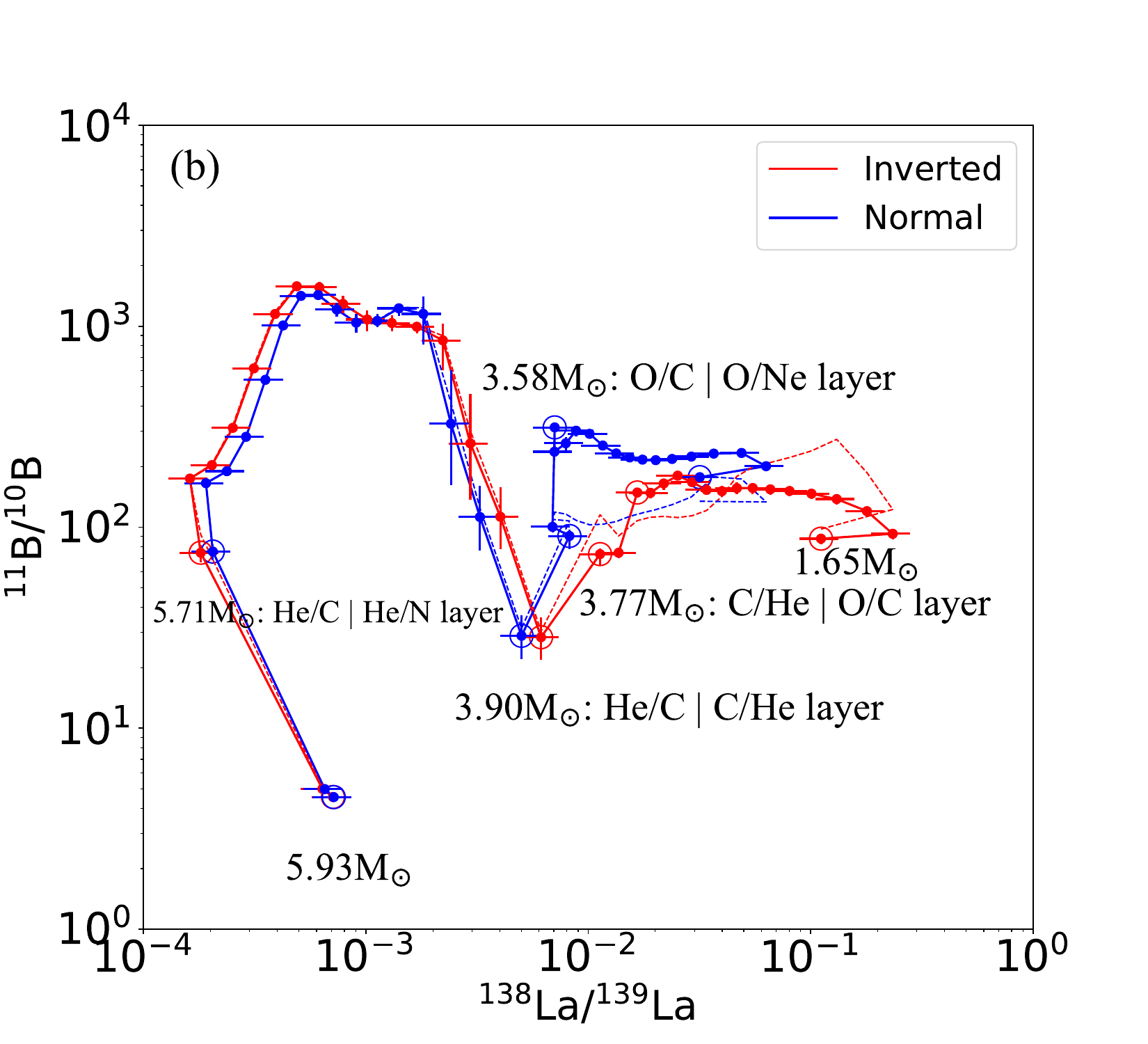}
    \caption{Calculated mass ratios (a) $\rm ^7Li/^6Li$ vs.  $\rm ^{138}La/^{139}La$ and (b) $\rm ^{11}B/^{10}B$ vs.  $\rm ^{138}La/^{139}La$. 
    The red and blue dots represent the results for the inverted and normal hierarchies at each position in the present SN model. 
    Each dot means a corresponding position of $M_r$ in Fig. \ref{c11_abun}. 
    The error bars represent the uncertainties of the abundance ratios arising from the uncertainties of $\nu$-induced reaction rates and $^{11}$C$(\alpha,p)^{14}$N reaction rate. 
    The points with circles indicate the particular locations in this SN model, i.e. the inner-most mass-coordinate at 1.65 $M_{\odot}$, 
    the boundaries between the O/Ne and O/C, O/C and C/He, C/He and He/C, He/C and He/N layers 
    at 3.58$M_{\odot}$, 3.77$M_{\odot}$, 3.90$M_{\odot}$, 5.71$M_{\odot}$,  respectively, 
    and the outermost mass coordinate at 5.93 $M_{\odot}$. 
    The dashed lines are the results using HW92. 
    %The value of solar system $\rm ^{138}La/^{139}La$, $^7$Li/$^6$Li and $^{11}$B/$^{10}$B are 8.96$\times 10^{-4}$, 
    %11.83$\pm$0.29 and 4.68$\pm$0.31 respectively \citep{Lodders:2009}. 
    }
    \label{2d_iso_ratio}
\end{figure}
\begin{figure}
    \centering
    \includegraphics[width=0.4\linewidth]{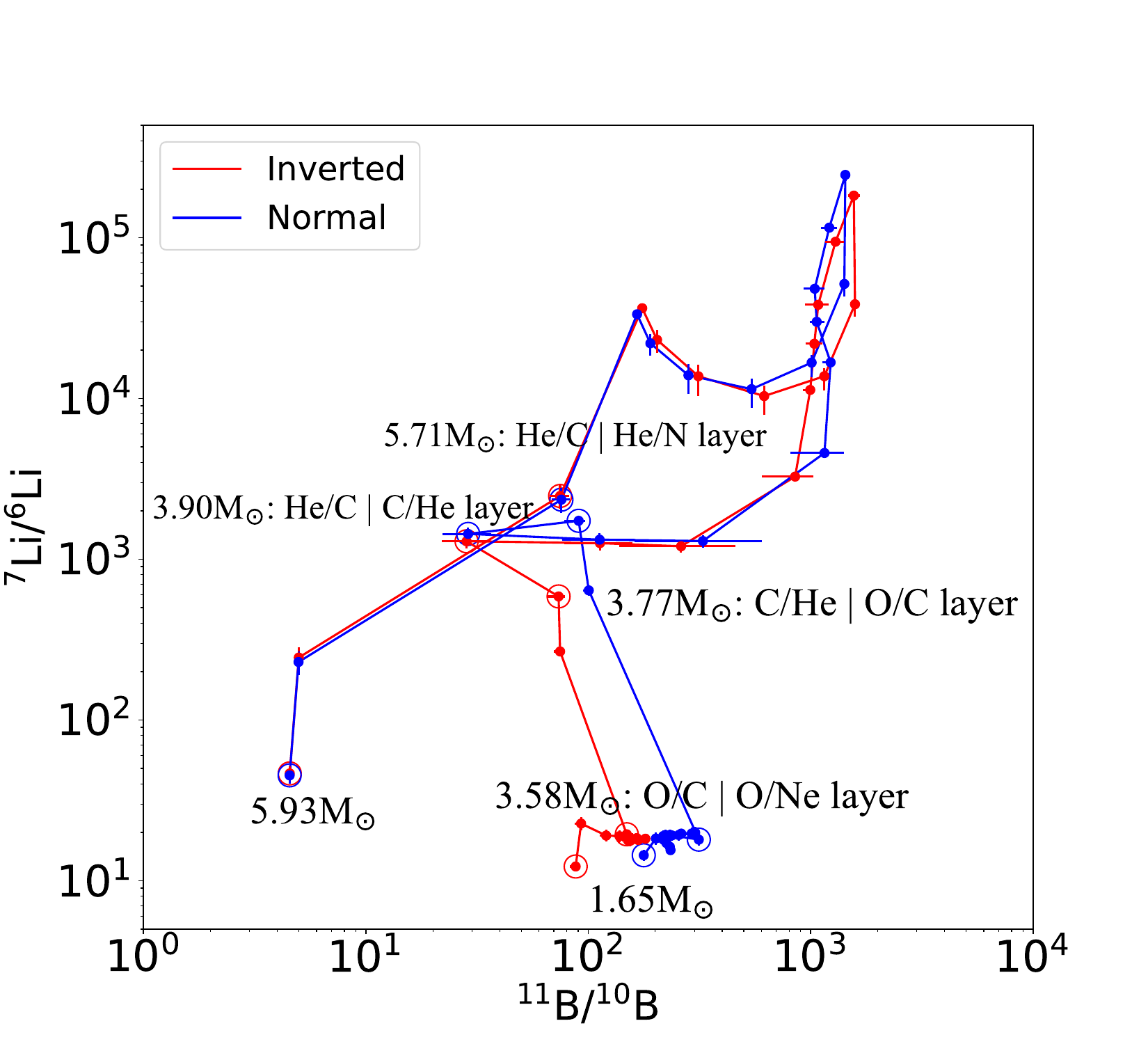}
    \caption{
    The same as Fig. \ref{2d_iso_ratio} but for
    $\rm (^7Li+{^7Be})/^6Li$ vs.  $\rm ({^{11}B}+{^{11}C})/(^{10}B+^{10}C)$. 
    }
    \label{fig:B_Li_ratio}
\end{figure}

\cite{Ko:2020rjq} have shown that the $^{138}$La/$^{11}$B ratio is sensitive to the mass hierarchy. 
We here extend this idea to measurements of $^{11}$B/$^{10}$B or $^7$Li/$^6$Li vs. $^{138}$La/$^{139}$La ratios in presolar grains originating from SNe. 
Type X SiC grains are robustly inferred to have condensed out of CCSN ejecta based on their multielement isotopic compositions \citep{Amari1992ApJ, Nittler1996ApJ,Hoppe1996,Liu2024SSRv}. 
Because SiC grains are condensed in the environment on condition C/O >1, 
they are formed by mixed materials from C-rich and Si-rich layers \citep{Lodders1995Metic, Hoppe2001}. 
$^{138}$La is mainly produced in the inner regions of the O/Ne layer through the C/He layers (region I, II, III), 
whereas $^{10}$B and $^{11}$B are produced in entire regions, 
particularly in the O/C and C/He layers (region II and III for $^{10}$B) and in the O/C and He/C layers (region II and IV for $^{11}$B). 
$^7$Li and $^6$Li are mainly produced in the He/C and C/He layers (region IV and III), respectively (see Fig. \ref{c11_abun}). 
To show the effects of neutrino mass hierarchies imprinted in nucleosynthetic observable, 
we take the isotopic abundance ratios of $^7$Li/$^6$Li, $^{11}$B/$^{10}$B, and $^{138}$La/$^{139}$La 
and display their correlations in Figs. \ref{2d_iso_ratio} and \ref{fig:B_Li_ratio}, 
where $^7$Li/$^6$Li and $^{11}$B/$^{10}$B stand for $\rm (^7Li+{^{7}Be})/^6Li$ and $\rm ({^{11}B}+{^{11}C})/({^{10}B}+{^{10}C})$ in nucleosynthesis calculations. 
Figure \ref{2d_iso_ratio} shows the $^7$Li/$^6$Li vs.  $\rm ^{138}La/^{139}La$ (panel (a)) and $^{11}$B/$^{10}$B vs.  $\rm ^{138}La/^{139}La$ (panel (b)) for the normal (blue solid lines) and inverted (red solid lines) hierarchies, respectively. 
Each point indicates the mass-fraction ratio averaged in a mass coordinate range of $\Delta M_r=$ 0.1224$M_{\odot}$. 
The points with circles indicate the particular locations in this model, i.e. the inner-most mass-coordinate at 1.65 $M_{\odot}$, 
the boundaries between the O/Ne and O/C, O/C and C/He, C/He and He/C, He/C and He/N layers 
at 3.58$M_{\odot}$, 3.77$M_{\odot}$, 3.90$M_{\odot}$, and 5.71$M_{\odot}$ respectively, 
and the outermost mass coordinate at 5.93 $M_{\odot}$. 

%An interesting consequence is inferred from Fig.\ref{2d_iso_ratio} that when we combine the $\rm ^{138}La/^{139}La$ ratio with $^7$Li/$^6$Li or $^{11}$B/$^{10}$B, 
The difference between the two neutrino hierarchies 
in $^7$Li/$^6$Li vs.  $\rm ^{138}La/^{139}La$ (Fig. \ref{2d_iso_ratio}(a)) and $^{11}$B/$^{10}$B vs.  $\rm ^{138}La/^{139}La$ (Fig. \ref{2d_iso_ratio}(b)) appears 
in the mass region between 3.58 and 3.90 $M_{\odot}$ 
which corresponds to the O/C and C/He layers. 
This is because $^7$Li and $^{11}$B abundances under the normal hierarchy are higher between 3.58 and 3.90 $M_{\odot}$ than those under the inverted hierarchy, 
while $^6$Li and $^{10}$B abundances under the normal hierarchy are lower than those under the inverted hierarchy (Figs. \ref{c11_abun}(b)-(d)). 
%$^7$Li/$^6$Li and $^{11}$B/$^{10}$B ratios are higher under normal hierarchy case at positions between 3.58 and 3.90 $M_{\odot}$, 
In Fig. \ref{2d_iso_ratio}(a), the two lines almost overlap in the inner layer at 1.63-3.58 $M_{\odot}$ for $^7$Li/$^6$Li vs. $^{138}$La/$^{139}$La. 
In contrast, for $^{11}$B/$^{10}$B vs. $^{138}$La/$^{139}$La in Fig. \ref{2d_iso_ratio}(b), 
the separation between the two lines is wider than that for $^7$Li/$^6$Li. 
This result suggests that combining boron and lanthanum may substantially constrain the neutrino mass hierarchy
rather than the combination of lithium and lanthanum. 
This separation arises from the remarkable difference of 
$^{138}$La at $M_r \leq 3.77M_{\odot}$ in Fig.\ref{c11_abun}(e) between the two hierarchies.  
The $\rm ^{138}La/^{139}La$ ratio is sensitive to the $\nu$-process 
because $^{138}$La is mainly produced in the $\nu$-induced reaction $\rm ^{138}Ba(\nu_e,e^-)^{138}La$ \citep{Heger2005PLB}.
On the other hand, the newer $\nu+^{16}$O rate from Su18 provides additional $^4$He in the nucleosynthesis at $M_r \leq 3.77M_{\odot}$, 
which separate the $^{11}$B/$^{10}$B ratio more significantly compared to the result using  HW92 (dashed lines in Fig. \ref{2d_iso_ratio}(b)). 
Such a result is valid even considering the uncertainties arising from nuclear cross sections in the model, as shown in the error bar in Fig. \ref{2d_iso_ratio}.

X SiC grains were found in the Murchison meteorite \citep{Amari1992ApJ, Nittler1996ApJ, Hoppe2000}. 
\cite{Hoppe2001} measured $^{11}$B/$^{10}$B in these X grains. 
The ratio is almost identical to the solar ratio within the uncertainty 
but a few orders of magnitude lower than the SN model prediction \citep{Woosley95}. 
To explain this fact, \cite{Hoppe2001} suggested three possible solutions:
1) changing initial neutrino temperature, 
2) mixing of layers with inhomogeneous isotopic abundance distribution, 
and 3) mixing of materials of C/O <1. 
As for 1), we mention that the lower neutrino temperature $T_{\nu_x}=6$ MeV rather than the temperature $T_{\nu_x}=8$ MeV \citep{Woosley95}, which \cite{Hoppe2001} referred to, 
has been adopted in the present calculation. 
The acceptable temperature range was found to be 4.8 MeV $\lesssim T_{\nu_x}\lesssim$ 6.6  MeV to reproduce the SN contribution to $^{11}$B in the framework of Galactic chemical evolution \citep{yoshida_constraining_2005}. 
As for 2) and 3), recent studies show that X grains form in CCSN ejecta at least after a few years but within thirty years \citep{Liu2018, Hartogh2022ApJ, Liu2024SSRv}. Therefore, there is a possibility that the CCSN ejecta from different layers could be mixed due to several hydrodynamic instabilities, 
micro-turbulence, or frictional ejection dynamics of materials in some stage before, 
during, or after the grain formation. 
Although clear evidence for such a mixing has not been observed in X grains, 
there is an indirect observational fact for the co-condensation of oxygen- and carbon-rich 
meteoritic stardust from nova outbursts \citep{Haenecour2019NatAs}, 
which suggests that the mixing process could operate in ejection process.

The average $^7$Li/$^6$Li ratio 11.83$\pm$0.29 and $^{11}$B/$^{10}$B ratio 4.68$\pm$0.31 in the twelve X grains were precisely measured by \cite{Fujiya_2011}. 
In Fig. \ref{fig:B_Li_ratio}, 
we present $^7$Li/$^6$Li vs.  $^{11}$B/$^{10}$B with the same setting as Fig. \ref{2d_iso_ratio}. 
\cite{Fujiya_2011} discussed the difference 
between their measurements and previous theoretical SN model calculations, e.g.,  \cite{Rauscher_2002}. 
Since X grains are known to form a few years after the supernova ejection  \citep{Liu2018, Hoppe2018GeCoA, Ott2019ApJ, Niculescu2022MNRAS}, 
significant effects could be caused by asteroidal or terrestrial contamination including the admixture of GCR components \citep{Liu2021ApJ, Gyngard2009}. 
By considering the Li/Si ratio 
and subtracting the GCR contribution, 
the measured ratio $^{11}$B/$^{10}$B=4.68$\pm$0.31 shows 
slight excess of $^{11}$B from the solar-system value $^{11}$B/$^{10}$B=4.03$^{+0.07}_{-0.02}$ \citep{zhai1996}. 
This suggests an admixture of other components such as the SN $\nu$-process product in these X grains. 
Figure. \ref{fig:B_Li_ratio} shows that none of the calculated $^7$Li/$^6$Li and $^{11}$B/$^{10}$B ratios for any SN layers can reproduce the values reported by \cite{Fujiya_2011}. 
This result supports the possibility that the observed lithium and boron may be subject to the mixing of a supernova ejecta and interstellar media whose composition is close to the solar materials or other  terrestrial contamination. 

%The recent study shows that X SiC grains may be formed at least two years after the supernova ejection \citep{Liu2018, Ott2019ApJ, Niculescu2022MNRAS} 
%, and the significant effects of laboratory contamination in presolar grains have been recognized by multielement study \citep{Liu2021ApJ.}

In Fig. \ref{2d_iso_ratio} \& \ref{fig:B_Li_ratio},
clear separation of the isotopic ratio divergence 
between normal and inverted hierarchies appears in the inner region $M_r\leq$ 3.90 $M_{\odot}$. 
%It has been confirmed that X SiC grains sample the material from the layer with C/O >1 \citep{Liu2024SSRv}. 
Thermodynamic calculations have shown that formation of SiC dust requires a condition with C/O>1 \citep{Lodders1995Metic}.
In our calculation, the C/He layer between 3.77 to 3.90 $M_{\odot}$ satisfies this condition of forming X grains.
Therefore, 
further search of X SiC presolar grains, especially of Li, B, and La isotopes is highly desirable in constraining the neutrino process and neutrino mass hierarchy. 
If other supernova presolar grains such as graphite, silicates, and oxides sampled materials from O-rich layers are found, 
our results would  provide possible constraint on the neutrino mass hierarchy.

\section{Conclusion}

We adopted new $\nu+^{16}$O and $\nu+^{20}$Ne reaction rates from a new shell-model 
and $\nu+^{138}$Ba rates from QRPA calculations in the neutrino-induced CCNS nucleosynthesis, 
including both collective and MSW oscillation effects. 
The newly added $\nu+^{16}$O and $\nu+^{20}$Ne reactions provide additional $^{6,7}$Li and $^4$He production. 
The $^4$He further modifies $^{11}$B abundance by $^{11}$C$(\alpha,p)^{14}$N reaction in the O/Ne and O/C and even C/He layers in Fig. \ref{c11_abun}. 
The collective oscillation effect enhances the high-energy tail of the $\nu_e$ spectrum in the inner layer under the inverted neutrino mass hierarchy by a factor of three. 
These two effects enlarge dramatically the separation of $^{138}$La/$^{139}$La and $^{11}$B/$^{10}$B 
ratios between the two hierarchies in $M_r < 3.90M_{\odot}$ 
as shown in Fig. \ref{2d_iso_ratio} (b). 
We also found the uncertainties of nuclear reactions and neutrino-induced reactions 
mostly come from $^{11}$C$(\alpha,p)^{14}$N, 
which affects the final yield of $^{11}$B in outer layer by a factor of two, 
but the uncertainty is negligible in the O/Ne through C/He layers ($1.7 \sim 3.9M_{\odot}$).  In our model, C/He layer (3.77 to 3.90 $M_{\odot}$) has C/O > 1, 
satisfies the X grains formation condition.
We propose that 
the isotopic ratios of $^{11}$B/$^{10}$B vs. $^{138}$La/$^{139}$La and $^7$Li/$^6$Li vs.  $^{138}$La/$^{139}$La in X grains could be 
a probe to the neutrino mass hierarchy and clarify the effect of neutrino flavor conversion
on the SN nucleosynthesis. 
%We also propose that these isotopic ratios in graphite, silicates, or oxides from O-rich layers would provide clearer constraints on the neutrino mass hierarchy. 
%when the isotopic ratios are measured together with of . 
%The result shows that the pair of $^{11}$B/$^{10}$B vs. $^{138}$La/$^{139}$La is more sensitive . 

%We emphasize that the split of isotopic ratios in Figs. \ref{2d_iso_ratio} \& \ref{fig:B_Li_ratio} between the two hierarchies may hint at clarifying the effect of neutrino flavor conversion on the SN nucleosynthesis.

%In these calculations, new $\rm ^{16}O+\nu$ and $^{20}$Ne+$\nu$ reaction rates from shell-model are adopted in nucleosynthesis, 
%which dramatically changes the production of $^{11}$B in the O/Ne layer. 
%Furthermore, the additional $\alpha$ particles produced from $\rm ^{16}O+\nu$ reaction destroys $^{11}$C by $(\alpha,p)$ reaction, 
%which reduces the difference of $^{11}$C abundance between the two hierarchies in the O/Ne layer. 

%Consequently, the final $^{11}$B abundance left in the SN ejecta shows only a little difference between two mass hierarchies. 
%Alternatively stated, the final total $^{11}$B yield in our SN model is not sensitive to neutrino hierarchy when $\rm ^{16}O+\nu$ is considered. 

%Many-body neutrino correlations have recently pointed out the collective neutrino oscillation  in addition to MSW, that neutrino flavour conversion is also discussed in literature (Wu2014,Yao2022,Fujimoto2022).
%Further 

\nolinenumbers
\section{Acknowledgment}
We gratefully acknowledge the referee of this paper for the kind and valuable suggestions. 
We also appreciate Hirokazu Sasaki and Motohiko Kusakabe for their in-depth discussions throughout this work. 
Yao was under the support of CSC scholarship from the Ministry of Education of China during his stay at the National Astronomical Observatory of Japan (NAOJ), where most of the numerical calculations for this paper were performed in the Center for Computational Astrophysics of NAOJ. 
This work was supported in part by the National Key R\&D Program of China (2022YFA1602401) and the National Natural Science Foundation of China (No. 12335009 \& 12435010). 
The work of M.K.C. is supported by the National Research Foundation of Korea (Grants No. NRF-2021R1A6A1A03043957 and No. NRF-2020R1A2C3006177).
\section{Appendix}
We show the isotope table, similar to \cite{meyer1995}, of our 20 $M_{\odot}$ supernova model. 
The masses are in units of $M_{\odot}$ for three cases without and with neutrino oscillation for normal and inverted mass hierarchies. 
These isotopic abundances are taken from our CCSN model calculation at 50 secs after the core-bounce.
\renewcommand{\arraystretch}{0.5}

\input{table1_together.tex}

\bibliography{Yao_bib}{}
\bibliographystyle{aasjournal}

\end{document}

%% file: table1_together.tex
\setlength{\tabcolsep}{2.5pt} 

\begin{longtable}{c|c|c c c c c c c c c}
\caption{Isotope source table for a 20$M_{\odot}$ core-collapse supernova.}
\label{nuclei_table}\\ 
\hline 
 & shell & Fe/Ni & Si/S  &O/Si  & O/Ne  & O/C  & C/He & He/C & He/N & Total \\ 
 & M(r) & 1.600 & 1.624 & 1.710 & 3.583 & 3.767 & 3.901 & 5.713 & 6.000 & 6.000 \\ 
 & $\Delta M$ & 0.012 & 0.024 & 0.086 & 1.873 & 0.184 & 0.134 & 1.812 & 0.287 & 6.000\\ 
\hline 
\endfirsthead\multicolumn{10}{c}{{\tablename} \thetable{}  -- Continued from previous page} \\ 
\hline
shell & & Fe/Ni & Si/S  &O/Si  & O/Ne  & O/C  & C/He & He/C &He/N & Total \\ 
M(r) & hierarchy & 1.600& 1.624 & 1.710 & 3.583 & 3.767 & 3.901 & 5.713 & 6.000 & 6.000 \\ 
$\Delta M$ & & 0.012 & 0.024 & 0.086 & 1.873 & 0.184 & 0.134 & 1.812 & 0.287 & 6.000\\ 
 \hline \endhead\multicolumn{10}{c}{{Continued on next page}} \\ 
\endfoot 
\hline 
\endlastfoot
\hline
				 & No-osci & 3.93E-08 & 1.57E-10 & 1.11E-10 & 3.64E-10 & 2.67E-11 & 1.43E-11 & 9.98E-09 & 2.18E-03 & 2.18E-03 \\ 
$\Delta M$($^{1}$H) & normal & 3.93E-08 & 1.96E-10 & 1.89E-10 & 6.32E-10 & 4.82E-11 & 2.58E-11 & 1.83E-08 & 2.18E-03 & 2.18E-03 \\ 
				 & inverted & 3.93E-08 & 1.79E-10 & 1.50E-10 & 6.66E-10 & 3.17E-11 & 2.24E-11 & 1.65E-08 & 2.18E-03 & 2.18E-03\\ 
\hline 
				 & No-osci & 4.52E-06 & 9.03E-09 & 1.82E-09 & 2.74E-10 & 2.20E-10 & 1.37E-10 & 7.77E-09 & 1.09E-02 & 6.96E-04 \\ 
$\Theta$($^{1}$H) & normal & 4.52E-06 & 1.13E-08 & 3.11E-09 & 4.76E-10 & 3.96E-10 & 2.47E-10 & 1.43E-08 & 1.09E-02 & 6.96E-04 \\ 
				 & inverted & 4.52E-06 & 1.03E-08 & 2.47E-09 & 5.01E-10 & 2.61E-10 & 2.15E-10 & 1.28E-08 & 1.09E-02 & 6.96E-04\\ 
\hline 
\hline
				 & No-osci & 1.83E-12 & 3.47E-12 & 9.11E-12 & 6.00E-11 & 9.58E-12 & 1.36E-11 & 5.98E-11 & 2.98E-09 & 3.14E-09 \\ 
$\Delta M$($^{2}$H) & normal & 1.83E-12 & 3.97E-12 & 6.75E-12 & 6.50E-12 & 3.90E-12 & 1.26E-11 & 5.82E-11 & 2.93E-09 & 3.02E-09 \\
\hline 
\end{longtable}